\documentclass[reprint,superscriptaddress,aps,twocolumn,showpacs]{revtex4-1}
\usepackage{amssymb}
\usepackage{amsmath}
\usepackage{graphicx}
\usepackage{multirow}
\usepackage{hyperref}

\allowdisplaybreaks

\begin{document}
\title{Meson cloud contributions to baryon axial form factors}

\author{X. Y. Liu}
\email[]{lxy\_gzu2005@126.com}
\author{K. Khosonthongkee}
\author{A. Limphirat}
\author{P. Suebka}
\author{Y. Yan}
\email[]{yupeng@sut.ac.th}
\affiliation{School of Physics, Institute of Science, Suranaree University of Technology,\\ Nakhon Ratchasima 30000, Thailand}
\affiliation{Thailand Center of Excellence in Physics (ThEP), Commission on Higher Education,\\ Bangkok 10400, Thailand}

\date{\today}

\begin{abstract}
The axial form factor as well as the axial charges and radii of octet $N$, $\Sigma$ and $\Xi$ baryons are studied in the perturbative chiral quark model with the quark wave functions predetermined by fitting the theoretical results of the proton charge form factor to experimental data. The theoretical results are found, based on the predetermined quark wave functions, in good agreement with experimental data and lattice values. This may indicate that the electric charge and axial charge distributions of the constituent quarks are the same. The study reveals that the meson cloud plays an important role in the axial charge of octet baryons, contributing 30\%--40\% to the total values, and strange sea quarks have a considerable contribution to the axial charges of the $\Sigma$ and $\Xi$.
\end{abstract}


\maketitle

\section{\label{sec:Intro}Introduction}

The form factors play an extremely important role in hadron physics since they supply necessary information on the internal structure and electroweak interaction properties. The $Q^2$ dependence of the electromagnetic and axial form factors of the nucleon have been studied in cloudy bag model~\cite{Theberge:1983,Thomas:1984}, lattice QCD~\cite{Liu24:1994,Lin49:2009,Yamazaki:2009,Syritsyn50:2010,Alexandrou25:2011,Bhattacharya50:2014} and other approaches~\cite{Cloet:2002,Matevosyan:2005,Schindler28:2007,Ramalho:2009,Erkol26:2011,Eichmann27:2012,Ramalho41:2011,*Ramalho42:2012,Ramalho43:2013}, in which the theoretical results are comparable with experimental data. The experimental and theoretical understanding of the electromagnetic and axial nucleon structure at low energy have been reviewed in Refs.~\cite{Bernard:2002,Arrington:2007}. In recent years, the hyperon axial charges, which are the axial form factors in zero recoil, have been predicted in lattice QCD~\cite{Lin43:2009,Erkol41:2010}, the chiral perturbation theory~\cite{Jiang30:2008,*Jiang30:2009}, and the relativistic constituent quark model (RCQM)~\cite{Choi431:2010}. However, there are few theoretical works on the $Q^2$ dependence of the axial form factor of hyperons, especially in the chiral quark model. This inspires us to study the axial form factors of octet baryons in the perturbative chiral quark model~(PCQM).

\begin{figure}[b!]
\begin{center}
\includegraphics[width=0.4\textwidth]{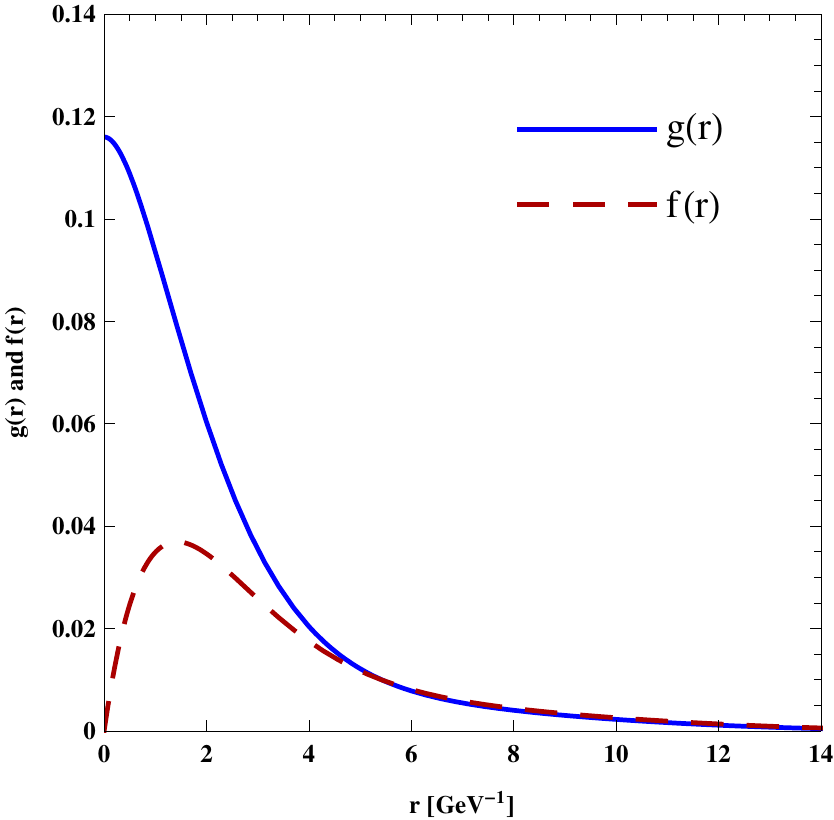}
\end{center}
\caption{Normalized radial wave functions of the valence quarks for the upper component $g(r)$ and the lower component $f(r)$ with the central values of the expansion coefficients, which are determined by fitting the theoretical results of the proton charge form factor to the experimental data~\cite{Liu1:2014}.}\label{fig:QWF}
\end{figure}

The PCQM~\cite{Lyubovitskij1:2001,Lyubovitskij3:2001,Lyubovitskij6:2002,Lyubovitskij2:2002,Pumsa-ard7:2003,Cheedket4:2004,Khosonthongkee10:2004,Dong8:2006,Dib9:2006,Faessler5:2008,Liu1:2014} is a powerful tool to study the baryon structure and properties in the low-energy particle physics. However, the previous work on the axial form factor of the nucleon~\cite{Khosonthongkee10:2004} shows that the PCQM theoretical result of the nucleon axial form factor is in good agreement with the experimental data only at very low momentum transfer $Q^2$, descending quickly with the momentum transfer $Q^2$ increasing. It is noted that a variational Gaussian ansatz has been employed for the quark wave functions~\cite{Khosonthongkee10:2004}. As we argue in Ref.~\cite{Liu1:2014}, the Gaussian-type quark wave functions of baryons lead to the theoretical predictions for the form factors of baryons consistent with experimental data only at very low momentum transfer $Q^2$. Furthermore, the more reasonable quark wave functions have been determined in Ref.~\cite{Liu1:2014} by fitting the PCQM theoretical result of the proton charge form factor to the experimental data, as shown in Fig.~\ref{fig:QWF}. In addition the $Q^2$ dependence of the theoretical electromagnetic form factors with the determined wave functions in the region $Q^2\leq1$ $\rm GeV^2$ is consistent with experimental data. More details could be found in Ref.~\cite{Liu1:2014}. In this work, we attempt to study the axial form factors of octet baryons in the PCQM with the determined wave functions in SU(3) and analyze the strangeness contributions to the axial form factors. We also predict the axial charges of light hyperons ($\Sigma$ and $\Xi$). There are no further parameters to be adjusted in the present work.

The paper is organized as follows. In Sec.~\ref{sec:PCQM&AFF}, we present the theoretical expressions of octet baryon axial form factors in the PCQM. The numerical results based on the predetermined quark wave functions and discussion are given in Sec.~\ref{sec:Results}.

\section{\label{sec:PCQM&AFF}Axial form factors in the PCQM}

In the framework of the PCQM, the axial form factors $G_A^B(Q^2)$ of octet baryons in the Breit frame are defined by
\begin{eqnarray}\label{eq:AFF}
\chi&_{B_{s'}}^\dag&\frac{\vec\sigma_B}{2}\chi_{B_s}G_A^B(Q^2)\nonumber\\
&=&\,^B\langle \phi_0|\sum_{n=0}^2 \frac{\mathrm{i}^n}{n!}\int \delta(t) d^4x d^4x_1\cdots d^4x_n e^{-\mathrm{i}q\cdot x}\nonumber\\
&&\times T[\mathcal{L}_I^W(x_1)\cdots \mathcal{L}_I^W(x_n)\vec A_3(x)]|\phi_0\rangle^B_c,
\end{eqnarray}
where the state vector $|\phi_0\rangle^B$ corresponds to the unperturbed three-quark states projected onto the respective baryon states, which are constructed in the
framework of the $SU(6)$ spin-flavor and $SU(3)$ color symmetry. The subscript \textit{c} in Eq.~(\ref{eq:AFF}) refers to contributions
from connected graphs only. $\chi_{B_s}$ and $\chi^\dag_{B_{s'}}$ are the baryon spin wave functions in the initial and final states, and $\vec\sigma_B$ is the baryon spin matrix. $G_A^B(Q^2)$ are the axial form factors of octet baryons with the squared momentum transfer $Q^2$.

The quark-meson interaction Lagrangian $\mathcal{L}_I^W(x)$ in Eq.~(\ref{eq:AFF}) takes the form
\begin{eqnarray}\label{eq:WT-int}
\mathcal{L}_I^W(x)&=&\frac{1}{2F}\partial_\mu\Phi_i(x)\bar{\psi}(x)\gamma^\mu\gamma^5
\lambda^i\psi(x)\nonumber\\
&&+\frac{f_{ijk}}{4F^2}\Phi_i(x)\partial_\mu\Phi_j(x)
\bar\psi(x)\gamma^\mu\lambda_k\psi(x),
\end{eqnarray}
where $F=88$ \textrm{MeV}; $\psi$ is the triplet of $u$, $d$ and $s$ quark fields; and $\Phi_i$ are the octet meson fields.\\
The axial-vector current $A^\mu_i$ in Eq.~(\ref{eq:AFF}) is given by
\begin{eqnarray}\label{eq:a-current}
A_i^\mu&=&
F\partial^\mu\Phi_i+\bar\psi\gamma^\mu\gamma^5\frac{\lambda_i}{2}\psi
-\frac{f_{ijk}}{2F}\bar\psi\gamma^\mu\lambda_j\psi\Phi_k\nonumber\\
&&+\bar\psi(\hat
Z-1)\gamma^\mu\gamma^5\frac{\lambda_i}{2}\psi+o(\Phi_i^2),
\end{eqnarray}
where the renormalization constant $\hat Z$ is determined by the nucleon charge conservation condition as
\begin{eqnarray}
\hat Z&=&1-\frac{3}{4(2\pi F)^2}\int_0^\infty dk k^4 F_{I}^2(k^2)\nonumber\\
&&\times\bigg[\frac{1}{\omega_\pi^3(k^2)}+\frac{2}{3\omega_K^3(k^2)}
+\frac{1}{9\omega_\eta^3(k^2)}\bigg],
\end{eqnarray}
with $\omega_\Phi(k^2)=\sqrt{M_\Phi^2+k^2}$ and the vertex function $F_{I}(k)$ for the $qq\Phi$ system taking the form
\begin{eqnarray}
F_{I}(k)&=&2\pi \int_0^\infty drr^2\int_0^\pi d\theta\sin\theta e^{i k r \cos\theta}\nonumber\\
&&\times[g(r)^2+f(r)^2\cos2\theta].
\end{eqnarray}

The ground-state quark wave function $u_0(\vec x)$ may, in general, be expressed as
\begin{equation}
u_0(\vec{x})=\left(\begin{array}{c}g(r)\\\large{i\vec{\sigma}\cdot\hat{x}f(r)}
\end{array}\right)\chi_s\chi_f\chi_c,
\end{equation}
where $\chi_s$, $\chi_f$ and $\chi_c$ are the spin, flavor and color quark wave functions, respectively. In the numerical analysis, we employ the radial quark wave functions $g(r)$ and $f(r)$ that have been extracted in Ref.~\cite{Liu1:2014} by fitting the theoretical results of the proton charge form factor to the experimental data. More information on the PCQM and quark wave functions can be found in Ref.~\cite{Liu1:2014}.

\begin{figure}
\begin{center}
\includegraphics[width=0.33\textwidth]{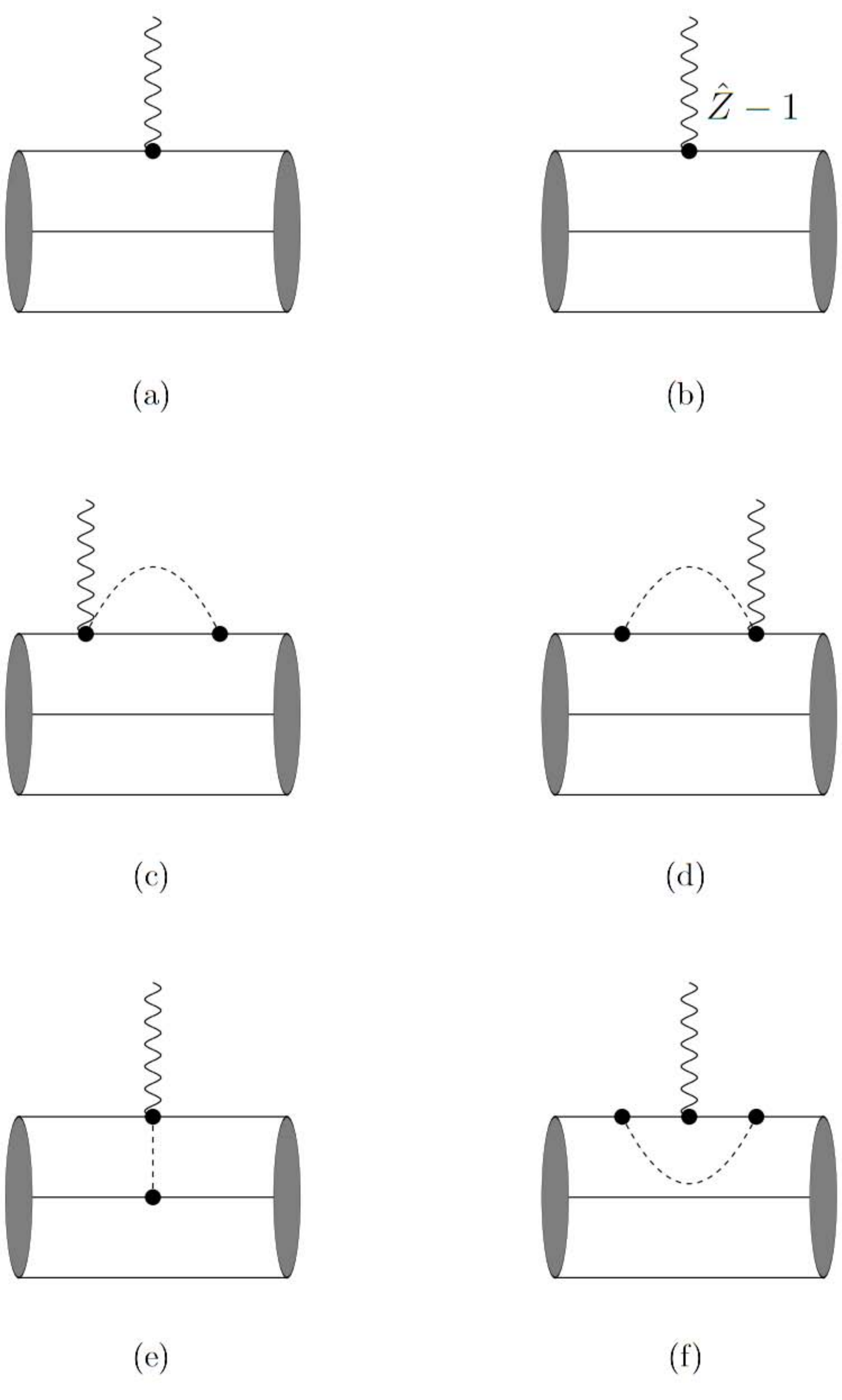}
\caption{\label{fig:AFF}Diagrams contributing to the axial form factor of octet baryons : 3q-core leading order (a), 3q-core counterterm (b), self-energy I (c), self-energy II (d), meson exchange (e), and vertex correction (f).}
\end{center}
\end{figure}

The Feynman diagrams contributing to the axial form factor of octet baryons in accordance with the $\mathcal{L}_I^W(x)$ in Eq.~(\ref{eq:WT-int}) and the $A^\mu_i$ in Eq.~(\ref{eq:a-current}) are shown in Fig.~\ref{fig:AFF}. The corresponding analytical expressions for the relevant diagrams are derived as follows:\\
\noindent(a) Three-quark core leading-order (LO) diagram:
\begin{eqnarray}\label{eq:LO}
G_A^B(Q^2)\big|_{LO}&=&c_1^B 2\pi\int_0^\infty drr^2 \int_0^\pi d\theta \sin\theta e^{i Q r \cos\theta}\nonumber\\
&&\times [g(r)^2+f(r)^2\cos(2\theta)].
\end{eqnarray}

\noindent(b) Three-quark core counterterm (CT) diagram:
\begin{equation}
G_A^B(Q^2)\big|_{CT}=(\hat Z-1)G_A^B(Q^2)\big|_{LO}.
\end{equation}

\noindent(c) Self-energy I (SE I) diagram:
\begin{eqnarray}\label{eq:SEI}
G_A^B(Q^2)\big|_{SE:I}&=&\frac{1}{2(2\pi F)^2}\int_0^\infty dkk^4 \int_{-1}^1 dx (1-x^2)\nonumber\\
&&\times\frac{F_{I}(k)F_{II}(k_-)}{\sqrt{k_-^2}}\bigg[\frac{c_1^B}{\omega_\pi^2(k^2)}
+\frac{c_2^B}{\omega_K^2(k^2)}\bigg].\nonumber\\
\end{eqnarray}
where $k_-=\sqrt{k^2+Q^2-2k\sqrt{Q^2}x}$, and the vertex function for the quark-pion-axial vector current $F_{II}(k)$ is given by
\begin{equation}
F_{II}(k)=-2i\pi \int_0^\infty drr^2 \int_0^\pi d\theta g(r)f(r)\sin2\theta e^{i k r \cos\theta}.
\end{equation}

\noindent(d) Self-energy II (SE II) diagram:
\begin{eqnarray}\label{eq:SEII}
G_A^B(Q^2)\big|_{SE:II}&=&\frac{1}{2(2\pi F)^2}\int_0^\infty dkk^4 \int_{-1}^1 dx (1-x^2)\nonumber\\
&&\times\frac{F_{I}(k)F_{II}(k_-)}{\sqrt{k_-^2}}\bigg[\frac{c_1^B}{\omega_\pi^2(k^2)}
+\frac{c_2^B}{\omega_K^2(k^2)}\bigg].\nonumber\\
\end{eqnarray}

\noindent(e) Exchange (EX) diagram:
\begin{eqnarray}
G_A^B(Q^2)\big|_{EX} &=& \frac{1}{4(2\pi F)^2}\int_0^\infty dkk^4 \int_{-1}^1 dx (1-x^2) \nonumber\\
&&\times\frac{F_{I}(k)F_{II}(k_-)}{\sqrt{k_-^2}}\bigg[\frac{c_3^B}{\omega_\pi^2(k^2)}
+\frac{c_4^B}{\omega_K^2(k^2)}\bigg].\nonumber\\
\end{eqnarray}

\noindent(f) Vertex-correction (VC) diagram:
\begin{eqnarray}\label{eq:VC}
G_A^{B}(Q^2)\big|_{VC}&=&\frac{1}{20(2\pi F)^2}\int_0^\infty dk k^4 F_{I}^2(k)\nonumber\\
&&\times\bigg[\frac{c_1^B}{\omega_\pi^3(k^2)}
+\frac{c_5^B}{\omega_\eta^3(k^2)}\bigg]\cdot G_A^{N}(Q^2)\big|_{LO}.\nonumber\\
\end{eqnarray}

The constants $c_i^B$ are given in Table~\ref{tab:c-constant}. It is noted that the constant $c_1^N=5/3$ in Eq.~(7) is determined by the spin and flavor of the three-quark core of the nucleon, namely the naive SU(3) quark model. In addition, Eqs.~(\ref{eq:SEI}) and~(\ref{eq:SEII}) present the same results for the diagrams (c) and (d) of Fig.~\ref{fig:AFF} based on the T symmetry.

\begin{table}[t!]
\caption{\label{tab:c-constant} The constants $c_i^B$ for the octet baryons axial form factors $G_A^B(Q^2)$.}
\begin{ruledtabular}
\begin{tabular}{@{}lccccc}
   \multicolumn{1}{l}{ }&
   \multicolumn{1}{c}{{}$\phantom{-}c_1$}&
   \multicolumn{1}{c}{{}$\phantom{-}c_2$}&
   \multicolumn{1}{c}{{}$c_3$}&
   \multicolumn{1}{c}{{}$\phantom{-}c_4$}&
   \multicolumn{1}{c}{{}$\phantom{-}c_5$}\\[2pt]
\hline\\[-13pt]
   $N$ & $\phantom{-}5/3$ & $\phantom{-}5/6$ & $8$ & $\phantom{-}0$ & $-5/9$\\[1pt]
   $\Sigma$ & $\phantom{-}4/3$ & $\phantom{-}2/3$ & $0$ & $\phantom{-}4$ & $-4/9$ \\[1pt]
   $\Xi$ & $-1/3$ & $-1/6$ & $0$ & $-4$ & $\phantom{-}1/9$\\
\end{tabular}
\end{ruledtabular}
\end{table}

\section{\label{sec:Results}Numerical results and discussion}

In this section, we present the axial charges and form factors of octet baryons with the determined quark wave functions~\cite{Liu1:2014}. The calculations are extended to the SU(3) flavor symmetry, including $\pi$, kaon and $\eta$-meson cloud contributions. Note that there are no further parameters in the following numerical calculations on the axial form factors of octet baryons.

\begin{table*}[t]
\caption{\label{tab:Axial charge}Numerical results for the octet baryon axial charges $g_A^B$, where the uncertainties are from the errors of the quark wave functions. The experimental data are taken from Ref.~\cite{PDG12:2014}, while the chiral extrapolation estimations of  lattice QCD results at the physical $m_\pi$ point are taken from Ref.~\cite{Lin43:2009}.}
\begin{ruledtabular}
\begin{tabular}{@{}lccccc}
   \multicolumn{1}{l}{ }&
   \multicolumn{1}{c}{$\phantom{-}$3q}&
   \multicolumn{1}{c}{$\phantom{-}$Meson loops}&
   \multicolumn{1}{c}{\multirow{2}{*}{$\phantom{-}$Total}}&
   \multicolumn{1}{c}{\multirow{2}{*}{$\phantom{-}$Lattice~\cite{Lin43:2009}}}&
   \multicolumn{1}{c}{\multirow{2}{*}{$\phantom{-}$Exp.~\cite{PDG12:2014}}}\\
   \multicolumn{1}{c}{ }&
   \multicolumn{1}{c}{$\phantom{-}$LO}&
   \multicolumn{1}{c}{$\phantom{-}$CT+SE+EX+VC}&
   \multicolumn{1}{c}{ }&
   \multicolumn{1}{c}{ }&
   \multicolumn{1}{c}{ }\\[2pt]
\hline\\[-13pt]
   $g_A^N$ & $\phantom{-}0.883$ & $\phantom{-}0.418$ & $\phantom{-}1.301\pm0.230$ & $\phantom{-}1.180\pm0.100$ & $1.272\pm0.002$\\[2pt]
   $g_A^\Sigma$ & $\phantom{-}0.707$ & $\phantom{-}0.220$ & $\phantom{-}0.927\pm0.132$ & $\phantom{-}0.900\pm0.096$& ---\\[2pt]
   $g_A^\Xi$ & $-0.177$ & $-0.106$ & $-0.283\pm0.033$ & $-0.277\pm0.034$& ---\\
\end{tabular}
\end{ruledtabular}
\end{table*}

The numerical results for the axial charges, which are the diagonal axial charges shown in Eq.~(\ref{eq:AFF}) with $\vec A_3$, are listed in Table~\ref{tab:Axial charge}. The uncertainties in the total values of the axial charges caused by the fitting errors of the quark wave functions~\cite{Liu1:2014} (the same hereinafter in Table~\ref{tab:Axial radii}) are estimated around 15\%. As shown in Table~\ref{tab:Axial charge}, the theoretical results reveal that the meson cloud plays an important role in the axial charge of octet baryons, contributing 30\%--40\% to the total values. Except for the $N$, there are no direct experimental data for the axial charge of the $\Sigma$ and $\Xi$, and thus we have the chiral extrapolation estimations of lattice QCD results at the physical $m_\pi$ point~\cite{Lin43:2009} compiled in the table for comparison. It is found that the theoretical $N$ axial charge is in good agreement with the experimental value~\cite{PDG12:2014}, and the work predictions on $\Sigma$ and $\Xi$ axial charges are consistent with the Lattice-QCD values~\cite{Lin43:2009}. In Ref.~\cite{Choi431:2010}, the axial charges of hyperons are evaluated in the RCQM without considering the chiral symmetry. Our tree-level (LO) results of the $\Sigma$ and $\Xi$ axial charges are comparable with the RCQM values while the meson loop diagrams contribute some correction to our tree-level results of $g_A^B$.

Listed in Table~\ref{tab:Axial radii} are the axial radii of octet baryons, which are derived by
\begin{equation}
\langle r^2_A\rangle_B=-6\frac{1}{g_A^B}\frac{dG_A^B(Q^2)}{dQ^2}|_{Q^2=0}.
\end{equation}
The nucleon axial radius $\langle r^2_A\rangle_N^{1/2}$ in Table~\ref{tab:Axial radii} is a little bit larger than the experimental value, and the predicted results for the $\langle r^2_A\rangle_\Sigma^{1/2}$ and $\langle r^2_A\rangle_\Xi^{1/2}$ are in the same order as $\langle r^2_A\rangle_N^{1/2}$ since our calculations are restricted to the SU(3) chiral symmetry. As discussed in Ref.~\cite{Khosonthongkee10:2004}, the contributions of excited-state quarks in loop diagrams generate some corrections to the $N$ axial form factor. The inclusion of the excited-state quarks in loop diagrams may be addressed in a future work.

\begin{table}[b!]
\caption{\label{tab:Axial radii}Numerical results for the octet baryon axial radii $\langle r^2_A\rangle_B^{1/2}$ (in units of $\rm fm$), where the uncertainties are from the errors of the quark wave functions. The experimental data are taken from Ref.~\cite{Bernard:2002}.}
\begin{ruledtabular}
\begin{tabular}{@{}lcc}
   \multicolumn{1}{l}{ }&
   \multicolumn{1}{c}{PCQM}&   \multicolumn{1}{c}{Exp.~\cite{Bernard:2002}}\\[2pt]
\hline\\[-13pt]
   $\langle r^2_A\rangle_N^{1/2}$ & $0.808\pm0.088$ & $0.639\pm0.010$\\[2pt]
   $\langle r^2_A\rangle_\Sigma^{1/2}$ & $0.832\pm0.089$& ---\\[2pt]
   $\langle r^2_A\rangle_\Xi^{1/2}$ & $0.780\pm0.087$& ---\\
\end{tabular}
\end{ruledtabular}
\end{table}

\begin{table}[b!]
\caption{\label{tab:Meson contri}Contribution of $\pi$, $K$ and $\eta$ mesons to the axial charges $g_A^B$.}
\begin{ruledtabular}
\begin{tabular}{@{}lccc}
   \multicolumn{1}{c}{ }&
   \multicolumn{1}{c}{ }&
   \multicolumn{1}{c}{Meson loops}&
   \multicolumn{1}{c}{ }\\
   \multicolumn{1}{c}{ }&
   \multicolumn{1}{c}{$\phantom{-}\pi$}&
   \multicolumn{1}{c}{$\phantom{-}K$}&
   \multicolumn{1}{c}{$\phantom{-}\eta$}\\[2pt]
\hline\\[-13pt]
   $g_A^N$ & $\phantom{-}0.375$ & $\phantom{-}0.045$ & $-0.002$ \\[2pt]
   $g_A^\Sigma$ & $\phantom{-}0.118$ & $\phantom{-}0.104$ & $-0.002$\\[2pt]
   $g_A^\Xi$ & $-0.030$ & $-0.077$ & $-0.001$\\
\end{tabular}
\end{ruledtabular}
\end{table}

\begin{table}[b!]
\caption{\label{tab:S-quark} Strange sea quark contributions of the individual loop diagrams of Fig.~\ref{fig:AFF} to the axial charges $g_A^B$.}
\begin{ruledtabular}
\begin{tabular}{@{}lcccc}
   \multicolumn{1}{l}{ }&
   \multicolumn{1}{c}{{}$\phantom{-}$CT}&
   \multicolumn{1}{c}{{}$\phantom{-}$SE}&
   \multicolumn{1}{c}{{}$\phantom{-}$EX}&
   \multicolumn{1}{c}{{}$\phantom{-}$VC}\\[2pt]
\hline\\[-13pt]
   $g_A^N$ & $-0.0136$ & $\phantom{-}0.0567$ & $\phantom{-}0\phantom{-}\phantom{-}\phantom{-}$ & $-0.0006$\\[1pt]
   $g_A^\Sigma$ & $-0.0109$ & $\phantom{-}0.0453$ & $\phantom{-}0.0680$ & $-0.0004$\\[1pt]
   $g_A^\Xi$ & $\phantom{-}0.0027$ & $-0.0113$ & $-0.0680$ & $\phantom{-}0.0001$\\
\end{tabular}
\end{ruledtabular}
\end{table}

Furthermore, we have studied the separate contribution of $\pi$, $K$ and $\eta$ mesons to the axial charges. As shown in Table~\ref{tab:Meson contri}, the $\pi$ meson contribution to the $N$ axial charge dominates over the ones from the $K$ and $\eta$ mesons, but the $K$ meson contributions to the $\Sigma$ and $\Xi$ axial charges are in the same order as the $\pi$ ones. It is noticed that the contribution from the $\eta$ meson is negligible. We also list in Table~\ref{tab:S-quark} the strange sea quark contributions ($K$ and $\eta$ meson clouds) of the individual loop diagrams as shown in Fig.~\ref{fig:AFF} to the axial charges $g_A^B$. Based on Eqs.~(\ref{eq:LO})--(\ref{eq:VC}), we may point out the fact that the $K$ meson contributes to the SE and EX diagrams while the $\eta$ meson participates in the VC process only. The results listed in Table~\ref{tab:S-quark} reveal that the strange sea quark contribution to the $N$ axial charge is caused mainly by the SE diagram, but to the $\Sigma$ and $\Xi$ axial charges both the SE and EX diagrams are important. As shown in the last column of Table~\ref{tab:S-quark}, the $\eta$ meson contribution is suppressed due to the weak coupling between the $s$ current quark and $\eta$ meson.

\begin{figure}[b!]
\begin{center}
\includegraphics[width=0.45\textwidth]{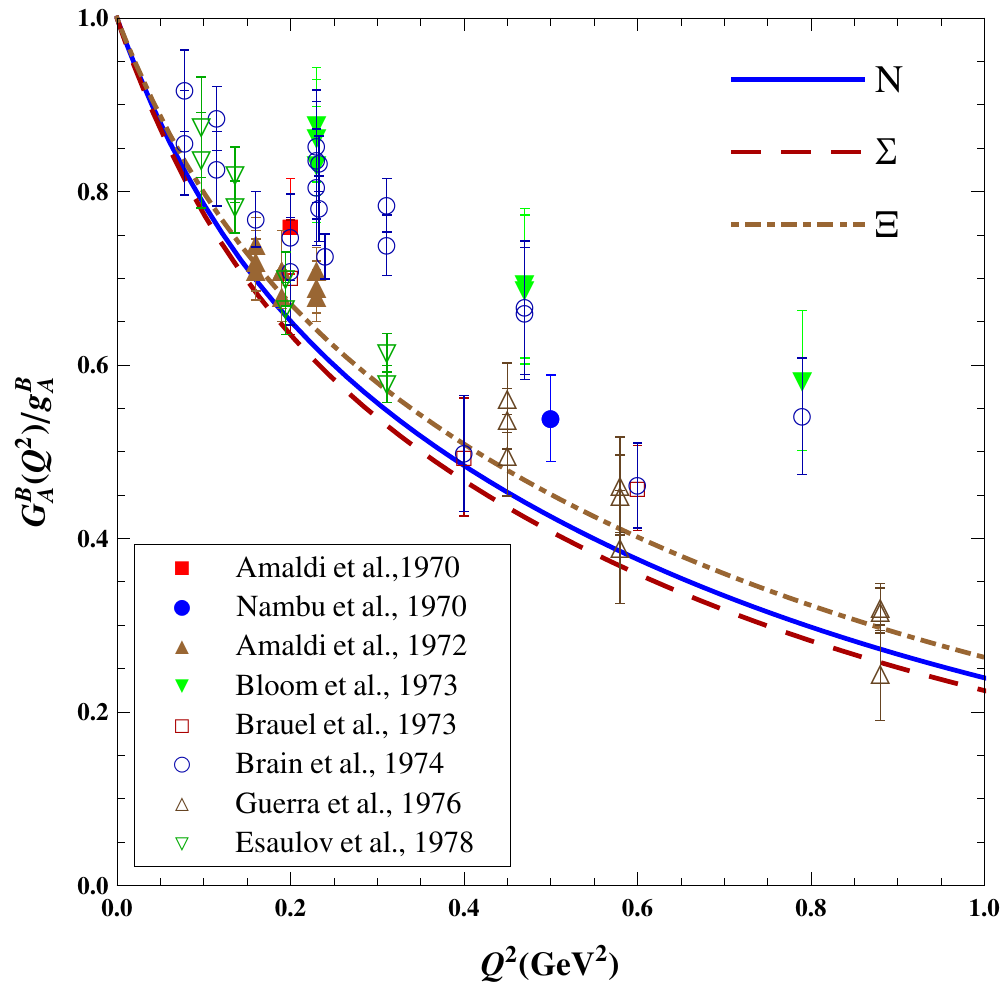}
\end{center}
\caption{\label{fig:GAB}Normalized axial form factors $G_A^B(Q^2)/g_A^B$ of octet baryons. The experimental data on nucleon axial form factor are taken from Refs.~\cite{Amaldi29:1970,Nambu30:1970,Amaldi31:1972,Bloom32:1973,Brauel33:1973,Read36:1974,Guerra35:1976,Esaulov34:1978}.}
\end{figure}

\begin{figure*}
\begin{center}
\includegraphics[width=0.3\textwidth]{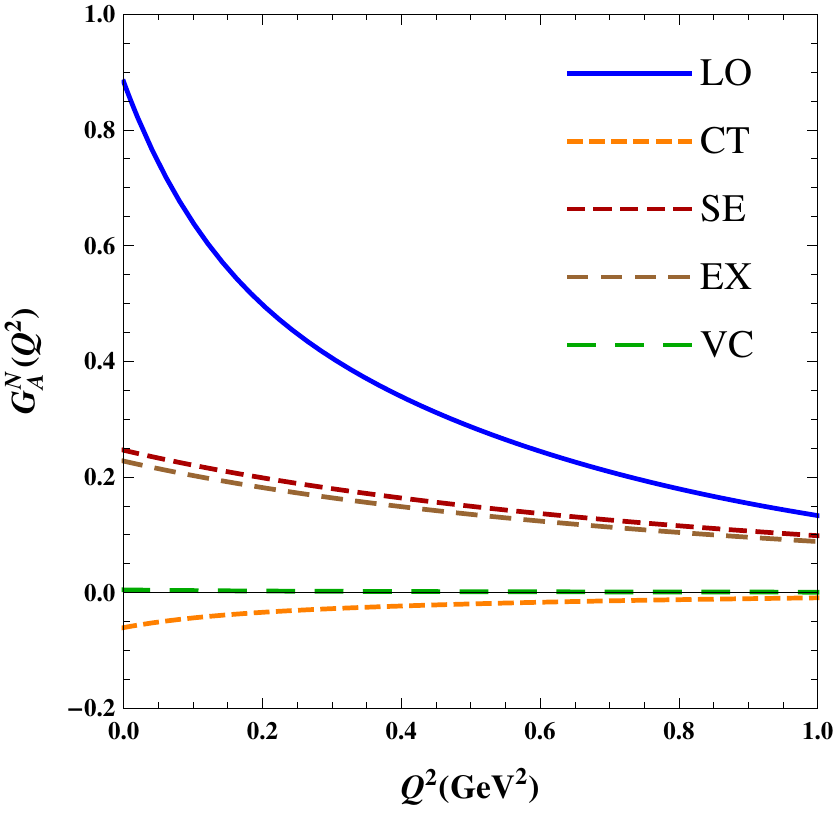}
\hspace {0.5cm}
\includegraphics[width=0.3\textwidth]{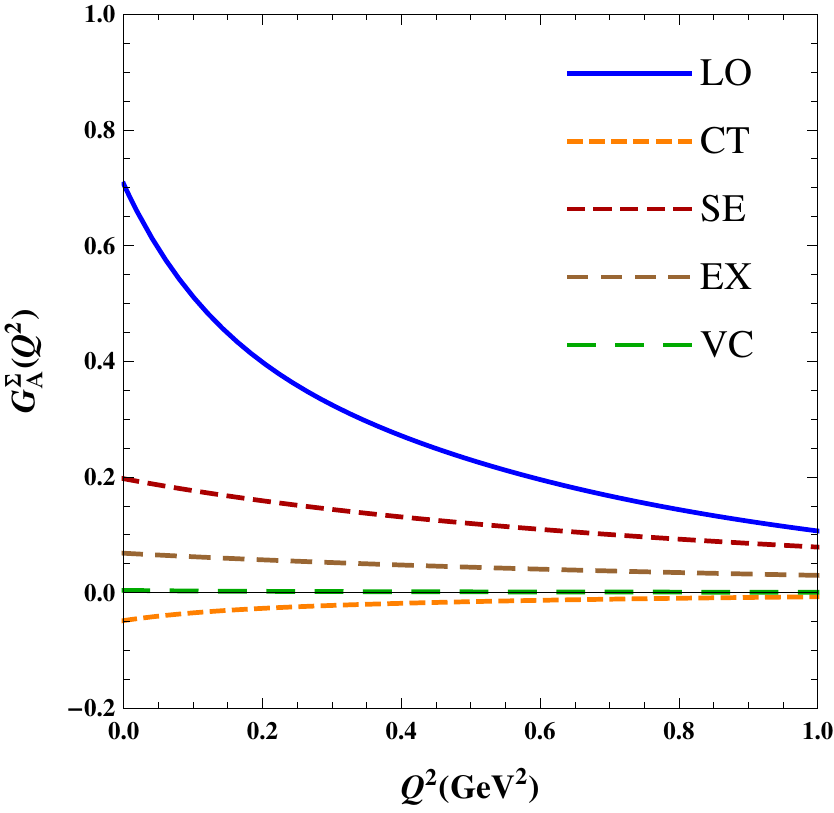}
\hspace {0.5cm}
\includegraphics[width=0.3\textwidth]{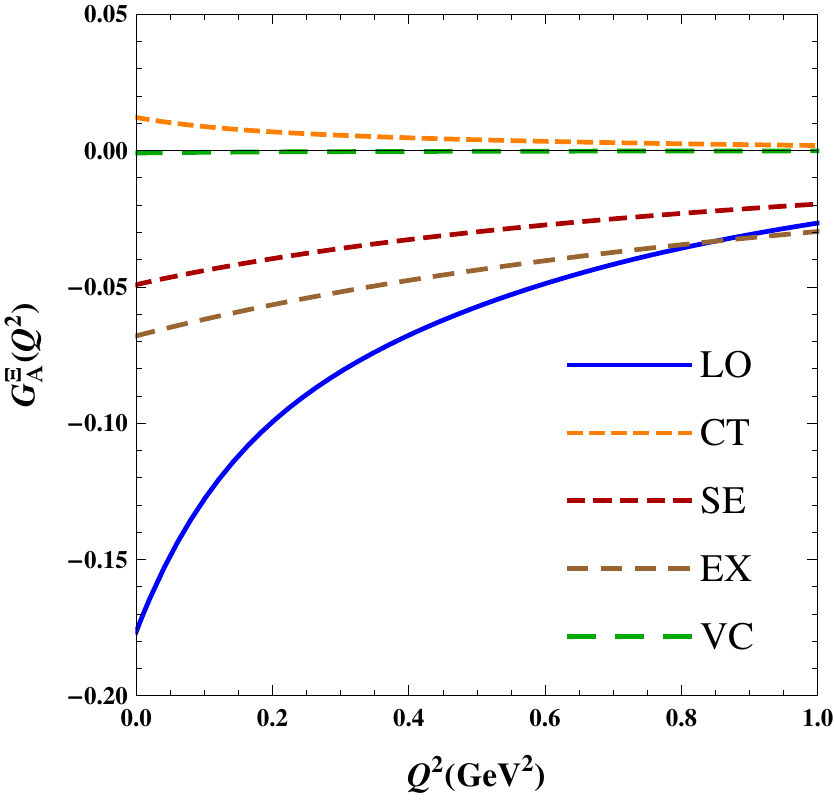}
\end{center}
\caption{\label{fig:GANSigmaXi}The individual contributions of the different diagrams of Fig.~\ref{fig:AFF} to the axial form factors of octet baryons (left panel for $N$, middle panel for $\Sigma$ and right panel for $\Xi$ ).}
\end{figure*}

We show the $Q^2$ dependence of the axial form factors of octet baryons in Fig.~\ref{fig:GAB}, which are normalized to 1 at zero recoil, with the experimental data on the nucleon axial form factor~\cite{Amaldi29:1970,Nambu30:1970,Amaldi31:1972,Bloom32:1973,Brauel33:1973,Read36:1974,Guerra35:1976,Esaulov34:1978} plotted as well. As shown in Fig.~\ref{fig:GAB}, the result for $G_A^N(Q^2)$ is close to the experimental data~\cite{Amaldi29:1970,Nambu30:1970,Amaldi31:1972,Bloom32:1973,Brauel33:1973,Read36:1974,Guerra35:1976,Esaulov34:1978} and the predicted results on $G_A^\Sigma(Q^2)$ and $G_A^\Xi(Q^2)$ show a similar $Q^2$ dependence based on the SU(3) symmetry. Considering the PQCM result of $g_A^N$ is $2.5\%$ larger than the experimental value, the non-normalized result for $G_A^N(Q^2)$ could be in better agreement with the experimental data. As expected, the theoretical axial form factors fall off smoothly when the momentum transfer $Q^2$ increases. The predetermined quark wave functions employed in the work take a form similar to Coulomb wave functions and have large values at small $r$ region as shown in Fig.~\ref{fig:QWF}, compared to the Gaussian-type wave functions employed in the previous work~\cite{Khosonthongkee10:2004}. This may be the main reason why the theoretical axial form factor evaluated with the predetermined wave functions is consistent with the experimental data especially at larger $Q^2$.

We present in Fig.~\ref{fig:GANSigmaXi} the contribution of various processes as shown in Fig.~\ref{fig:AFF} to the axial form factors of octet baryons. It is found that the sea quark or meson cloud contributes to axial form factors mainly through the SE and EX diagrams. The LO diagram results in a dipolelike axial form factor while the meson cloud leads to a flat contribution to the axial form factor. The flat contribution indicates that the sea quarks distribute mainly in a very small region, which is rather surprising and needs to be further studied.

In summary, one may conclude that the fact that the theoretical results of the axial form factors and axial charges agree well with experimental data and lattice QCD values, with the predetermined quark core wave functions in the electromagnetic sector, may indicate that the electric charge and axial charge distributions of the constituent quarks are the same. The study reveals that the meson cloud plays an important role in the axial charge of octet baryons, contributing 30\%--40\% to the total values, and strange sea quarks have a considerable contribution to the axial charge of the $\Sigma$ and $\Xi$.

The center-of-mass correction has been considered in relativistic quark models in Refs.~\cite{Lu:1998,Dong:1999,Tursunov:2014}. The nucleon mass is very sensitive to the center-of-mass effect, decreasing some 40\%~\cite{Tursunov:2014} while the theoretical results in Ref.~\cite{Lu:1998} reveal that the center-of-mass correction reduces the magnetic moments of the nucleon by about 10\%. Therefore, it is necessary to investigate the center-of-mass correction in the PCQM in our future work.

\begin{acknowledgments}
This work is supported by Suranaree University of Technology and CHE-NRU Project No. NV28/2556.
\end{acknowledgments}

\bibliography{Refs}

\end{document}